\documentclass[aps,twocolumn,amssymb,amsmath,superscriptaddress]{revtex4-1}

\usepackage{bm}
\usepackage{color}
\usepackage{graphicx}
\usepackage{dcolumn}
\usepackage{graphicx}
\usepackage[colorlinks=true, letterpaper=true, pdfstartview=FitV, linkcolor=blue, citecolor=blue, urlcolor=blue]{hyperref}
\usepackage[normalem]{ulem}
\usepackage{amsmath}
\usepackage{epstopdf}
\usepackage{multirow}
\usepackage{slashed}
\usepackage{float}

\setcitestyle{square}

\makeatletter
\renewcommand\@biblabel[1]{#1.}
\makeatother

\begin{document}

\title{Intrinsic and extrinsic anomalous transport properties of Heusler ferromagnets Fe$_2$CoAl and Fe$_2$NiAl from first principles}

\author{Xiuxian Yang}
\affiliation{Centre for Quantum Physics, Key Laboratory of Advanced Optoelectronic Quantum Architecture and Measurement (MOE),School of Physics, Beijing Institute of Technology, Beijing 100081, China}
\affiliation{Beijing Key Lab of Nanophotonics and Ultrafine Optoelectronic Systems, School of Physics, Beijing Institute of Technology, Beijing 100081, China}    

\author{Wanxiang Feng}
\email{wxfeng@bit.edu.cn}
\affiliation{Centre for Quantum Physics, Key Laboratory of Advanced Optoelectronic Quantum Architecture and Measurement (MOE),School of Physics, Beijing Institute of Technology, Beijing 100081, China}
\affiliation{Beijing Key Lab of Nanophotonics and Ultrafine Optoelectronic Systems, School of Physics, Beijing Institute of Technology, Beijing 100081, China}    

\author{Xiao-Ping Li}
\affiliation{School of Physical Science and Technology, Inner Mongolia University, Hohhot 010021, China}

\author{Gui-Bin Liu}
\affiliation{Centre for Quantum Physics, Key Laboratory of Advanced Optoelectronic Quantum Architecture and Measurement (MOE),School of Physics, Beijing Institute of Technology, Beijing 100081, China}
\affiliation{Beijing Key Lab of Nanophotonics and Ultrafine Optoelectronic Systems, School of Physics, Beijing Institute of Technology, Beijing 100081, China}    

\author{Yuriy Mokrousov}
\affiliation{Institute of Physics, Johannes Gutenberg University Mainz, 55099 Mainz, Germany}
\affiliation{Peter Gr\"unberg Institut and Institute for Advanced Simulation, Forschungszentrum J\"ulich and JARA, 52425 J\"ulich, Germany}

\author{Yugui Yao}
\email{ygyao@bit.edu.cn}
\affiliation{Centre for Quantum Physics, Key Laboratory of Advanced Optoelectronic Quantum Architecture and Measurement (MOE),School of Physics, Beijing Institute of Technology, Beijing 100081, China}
\affiliation{Beijing Key Lab of Nanophotonics and Ultrafine Optoelectronic Systems, School of Physics, Beijing Institute of Technology, Beijing 100081, China}

\date{\today}

\begin{abstract}
Recently, Heusler ferromagnets have been found to exhibit unconventional anomalous electric, thermal, and thermoelectric transport properties. In this study, we employed first-principles density functional theory calculations to systematically investigate both intrinsic and extrinsic contributions to the anomalous Hall effect (AHE), anomalous Nernst effect (ANE), and anomalous thermal Hall effect (ATHE) in two Heusler ferromagnets: Fe$_2$CoAl and Fe$_2$NiAl.  Our analysis reveals that the extrinsic mechanism originating from disorder dominates the AHE and ATHE in Fe$_2$CoAl , primarily due to the steep band dispersions across the Fermi energy and corresponding high longitudinal electronic conductivity. Conversely, the intrinsic Berry phase mechanism, physically linked to nearly flat bands around the Fermi energy and gapped by spin-orbit interaction band crossings, governs the AHE and ATHE in Fe$_2$NiAl. With respect to ANE, both intrinsic and extrinsic mechanisms are competing  in Fe$_2$CoAl as well as in Fe$_2$NiAl.  Furthermore, Fe$_2$CoAl and Fe$_2$NiAl exhibit tunable and remarkably pronounced anomalous transport properties.  For instance, the anomalous Nernst and anomalous thermal Hall conductivities in Fe$_2$NiAl attain giant values of 8.29 A/Km and 1.19 W/Km, respectively, at room temperature.  To provide a useful comparison, we also thoroughly investigated the anomalous transport properties of Co$_2$MnGa.  Our findings suggest that Heusler ferromagnets Fe$_2$CoAl and Fe$_2$NiAl are promising candidates for spintronics and spin-caloritronics applications.
\end{abstract}

\maketitle

\section{Introduction}\label{intro}
The anomalous Hall effect (AHE), a transverse voltage drop induced by a longitudinal charge current in the absence of an external magnetic field, is one of the most fundamental manifestations of magnetism and it has been exploited extensively for various applications in spintronics~\cite{Nagaosa2010}.  In addition, the anomalous Nernst effect (ANE)~\cite{Xiao2006} and anomalous thermal Hall effect (ATHE)~\cite{QinTao2011}---the thermoelectric and thermal analogues of the AHE---are two other important anomalous transport phenomena with particularly exciting prospects in spin-caloritronics~\cite{Bauer2012,Boona2014}.  It is well known that the origin of anomalous transport phenomena in diverse magnetic materials is fairly complicated but can be generally separated into intrinsic and extrinsic mechanisms.  The intrinsic part can be described well in terms of Berry phase theory in perfect crystals, which is independent of disorder details rooting in geometry of the electronic structure~\cite{Karplus1954,Sundaram1999,Onoda2002,Jungwirth2002}.  A large number of experimental and theoretical works have reported that the intrinsic mechanism dominates in many ferromagnetic and antiferromagnetic materials, in which large intrinsic anomalous Hall, anomalous Nernst, and anomalous thermal Hall conductivities (AHC, ANC, and ATHC) are expected owing to the emergence of large Berry curvature in momentum space~\cite{ZhongF2003,Yao2004,Ajaya2016,Suzuki2016,WangQi2018,Enke2018,Manna2018,Zhou2019,ZhouXD2020,Ilya2019,Akito2018,Jonathan2020,LiPG2022}.  For example, Heusler ferromagnets Co$_2$MnGa~\cite{Ilya2019,Akito2018} and Co$_2$MnAl~\cite{Jonathan2020,LiPG2022} show the large AHC on the order of 10$^3$ S/cm, which is recognized to be of an intrinsic origin.

On the other hand, the extrinsic mechanism includes the side jump~\cite{Berger1970} and skew scattering~\cite{Smit1955,Smit1958} contributions, both of which come from spin-orbit mediated electron scattering off disorder.  Recent experiments reveal that the extrinsic mechanism often cannot be ignored: for example, the AHC dominated by the extrinsic mechanism has been found in relatively ``dirty" kagome magnets~\cite{YangSY2020_2,Singh2021} and magnetic van der Waals materials~\cite{HuangM2021}, with AHC values  reaching up to the order of 10$^4$ $\sim$ 10$^5$ S/cm, overwhelming the intrinsic contributions by far.  Most of previous theoretical works focusing on the extrinsic mechanism adopt the effective models which are applicable in simple Weyl and Dirac semimetals~\cite{Burkov2014,Shapourian2016,AdoIA2017,Keser2019,Papaj2021}, but which are not able to match the complexity of band structures in real materials.  At the level of first-principles calculations, the extrinsic mechanism was paid much less attention as opposed to intrinsic mechanism addressed in numerous works. This is quite surprising, since a unified treatment of both the intrinsic and extrinsic contributions to the anomalous transport in realistic magnetic materials is the key to practical implementations of various concepts in spintronics and spin-caloritronics.
 
The family of Heusler compounds is considerably large comprising more than 1,000 members, which can reside in paramagnetic, ferromagnetic, and antiferromagnetic state, and display fascinating anomalous electric, thermal, and thermoelectric transport properties.  The AHE has been reported to be intrinsic in a number of Heusler compounds~\cite{Ilya2019,Akito2018,Jonathan2020,LiPG2022}, but can also show a transition from the intrinsic to extrinsic mechanism, e.g., in PrAlGe$_{1-x}$Si$_x$ with increasing the alloy ratio $x$~\cite{YangHY2020}.  Currently, a systematic investigation of the intrinsic and extrinsic anomalous transport properties in Heusler compounds is   lacking. Among Heusler compounds, Fe$_2$CoAl and  Fe$_2$NiAl~\cite{Buschow1983,SZYMANSKI2000,Sherder2012,Vishal2013,MENUSHENKOV2015,Saito2018,AHMAD2021,Ahmad2020EPJB,Felix2021} share the same crystal and magnetic structure with Co$_2$MnGa and Co$_2$MnAl, but their Curie temperatures ($830\sim1010$ K)~\cite{Saito2018,AHMAD2021} are higher  (690~K for Co$_2$MnGa ~\cite{Akito2018,Ilya2019} and  726~K for Co$_2$MnAl~\cite{Umetsu2008}).

In this work, utilizing first-principles density functional theory calculations, we systematically investigate the intrinsic and extrinsic contributions to the AHE, ANE, and ATHE in ferromagnetic Heusler compounds Fe$_2$CoAl and Fe$_2$NiAl.  First, the structure of the conductivity tensor is determined from the symmetry group analysis.  Then, different sources of anomalous transport properties, including intrinsic, side jump, and skew scattering contributions, are calculated individually and their competition are clearly explored.  In the case of AHE and ATHE, we show that the extrinsic mechanism dominates in Fe$_2$CoAl, while the intrinsic contributions play a crucial role in Fe$_2$NiAl.  For the ANE, the intrinsic and extrinsic mechanisms are highly competitive in both Fe$_2$CoAl and Fe$_2$NiAl.  By analyzing longitudinal electronic conductivity, electronic structure and Berry curvature distribution,  the underlying physics of the extrinsic and intrinsic mechanism-dominated anomalous transport properties in Fe$_2$CoAl and Fe$_2$NiAl is elucidated. We contrast our findings to the case of famous Heusler ferromagnet Co$_2$MnGa.  We also find that the ratio of intrinsic and extrinsic contributions can be efficiently tuned, and  total AHC, ANC, and ATHC can be significantly enhanced by doping.  For example, the total ANC and ATHC in Fe$_2$NiAl can be brought to reach gigantic values of 8.29 A/Km and 1.19 W/Km at room temperature, respectively.  Our findings promote two outstanding host materials for realizing exciting spintronics- and spincaloritronics-based applications  in information processing and energy conversion.

\section{Theory and computational details}\label{method}

The anomalous transport coefficients can be assessed from expressions derived within the Landauer-B\"uttiker formalism~\cite{ashcroft1976solid,Houten1992,behnia2015fundamentals}:
\begin{equation}\label{eq:LB}
	R^{(n)}_{ij}=\int^\infty_{-\infty}(E-\mu)^n\left(-\frac{\partial f}{\partial E}\right)\sigma_{ij}(E)\textnormal{d}E,
\end{equation}
where $\mu$ is chemical potential, $f =1/[\text{exp}((E-\mu)/k_{B}T)+1]$ is Fermi-Dirac distribution function, and $\sigma_{ij}$ is the AHC.  Then, the temperature-dependent ANC ($\alpha_{ij}$) and ATHC ($\kappa_{ij}$) respectively read
\begin{eqnarray}
	\alpha_{ij} &=& -R^{(1)}_{ij}/eT, \\  \label{eq:ANC}
	\kappa_{ij} &=& R^{(2)}_{ij}/e^2T. \label{eq:ATHC}
\end{eqnarray}
It is clear that the AHC is the key ingredient to capture other anomalous transport properties in target magnetic materials.

According to Kubo linear-response formalism~\cite{Kubo1957}, the AHC can be separated into the Fermi surface ($\sigma_{ij}^{\textnormal{I}}$) and Fermi sea ($\sigma_{ij}^{\textnormal{II}}$) terms~\cite{Czaja2014}:
\begin{eqnarray}\label{eq:int_1}
\sigma_{ij}^{\textnormal{I}}&=&-\frac{e^2\hbar}{2\pi}\int\frac{\textnormal{d}^3k}{(2\pi)^3}\sum_{m\neq n}{\rm Im}[v_{mn}^i(\textbf{k})v_{nm}^j(\textbf{k})] \nonumber\\
&=&\frac{(E_{m\textbf{k}}-E_{n\textbf{k}})\Gamma}{\left[(E_f-E_{m\textbf{k}})^2+\Gamma^2\right]\left[(E_f-E_{n\textbf{k}})^2+\Gamma^2\right]},
\end{eqnarray}
and
\begin{eqnarray}\label{eq:int_2}
\sigma_{ij}^{\textnormal{II}}&=&\frac{e^2\hbar}{\pi}\int\frac{\textnormal{d}^3k}{(2\pi)^3}\sum_{m\neq n}{\rm Im}[v_{mn}^i(\textbf{k})v_{nm}^j(\textbf{k})] \nonumber\\
&=&\left\{\frac{\Gamma}{(E_{m\textbf{k}}-E_{n\textbf{k}})[(E_f-E_{m\textbf{k}})^2+\Gamma^2]}\right.  \nonumber\\
&&\left. -\frac{1}{(E_{m\textbf{k}}-E_{n\textbf{k}})^2}{\rm Im}\left[{\rm In}\frac{E_f-E_m\textbf{k}+\textbf{i}\Gamma}{E_f-E_n\textbf{k}+\textbf{i}\Gamma}\right]\right\},
\end{eqnarray}
where $v^{i,j}$ are the velocity operators along the directions $\{i,j\}\in\{x,y,z\}$ with $x,y,z$ as Cartesian coordinates, $E_{n\textbf{k}}$ is  the energy eigenvalue of band  $n$ and Bloch momentum $\textbf{k}$, $E_f$ is the Fermi energy, and $\Gamma$ is a constant smearing parameter (0 $\sim$ 0.05 eV), respectively.  The sum of Eqs.~\eqref{eq:int_1} and~\eqref{eq:int_2} is the intrinsic AHC ($\sigma^\textnormal{int}_{ij}$), which converges to the well-known Berry curvature expression in the clean limit ($\Gamma\rightarrow0$)~\cite{Yao2004}:
\begin{equation}\label{eq:IAHC} 
	\sigma^\textnormal{int}_{ij} = e^{2}\hbar \int\frac{\textnormal{d}^{3}k}{(2\pi)^{3}}\sum^\textnormal{occ}_{n,m \neq n}\frac{2\mathrm{Im}\left[ v^i_{mn}(\textbf{k})v^j_{nm}(\textbf{k})\right]}{\left(E_{m\textbf{k}}-E_{n\textbf{k}}\right)^{2}}.
\end{equation}
While physically, a constant $\Gamma$ parameter mimics the effect of constant smearing (CS) in the sense that all electronic states acquire the same  finite lifetime, more intricate scattering mechanisms can be taken care of by going beyond the CS model. For example, the scattering-originated contributions (side jump and skew scattering) to the AHE can be taken into account by incorporating a short-range Gaussian disorder potential~\cite{Czaja2014}.

In the Gaussian disorder (GD) model, the impurity potential is represented by a set of delta functions located at random positions $\textbf{R}_{i}$:
\begin{equation}\label{eq:potential} 
V=U\sum_{i}^{N}\delta(\hat{\textbf{r}}-\textbf{R}_{i}).
\end{equation}
Here, $U$ is a measure for the scattering strength, and $N$ is the number of impurities ($n_{i}=N/V$ is the impurity concentration, where $V$ is the volume of the cell).  Since the exact distribution of impurities over the crystal can never be known, it is therefore feasible to calculate the actual Green function by taking the configurational average over all possible distributions of $N$ impurities, implying $G\equiv \left\langle G\right\rangle_{c}$.  The full Green's functions in the retarded ($R$) and advanced ($A$) forms are expressed as:
\begin{eqnarray}\label{eq:Green}
	G^{R}(E,\textbf{k})&=&[E-H(\textbf{k})-\Sigma(E,\textbf{k})]^{-1}, \\
	G^{A}(E,\textbf{k})&=&G^{R}(E,\textbf{k})^\dagger.
\end{eqnarray}
Here, $H(\textbf{k})$ is the Hamiltonian in the basis of Wannier functions. The self-energy $\Sigma(E,\textbf{k})$, accounting for the effect of electron scattering off disorder, can be written as follows, truncated to the lowest order~\cite{Czaja2014}:
\begin{equation}\label{eq:sigma}
	\Sigma(E,\textbf{k})=\mathcal{V}\int\frac{\textnormal{d}^3k'}{(2\pi)^3}O_{\textbf{kk}'}G_0(E,\textbf{k}')O_{\textbf{k}'\textbf{k}},
\end{equation}
where $\mathcal{V}=U^2n_\textnormal{i}$ stands for the disorder parameter, $O_{\textbf{kk}'}$ is the overlap matrix for the eigenstates at different wavevectors, and $G_0(E,\textbf{k}')=[E-H(\textbf{k}')]^{-1}$ is the unperturbed Green's functions.

The total AHC can be expressed in terms of the retarded and advanced full Green functions ($G^{R/A}$) that incorporate the effect of disorder~\cite{Czaja2014}:
\begin{eqnarray}\label{eq:ext_1}
\sigma_{ij}^\textnormal{I}&=&\frac{e^2\hbar}{4\pi}\int\frac{\textnormal{d}^3k}{(2\pi)^3}{\rm Tr}[\boldsymbol{\Gamma}^i (E_f, \textbf{k})G^R(E_f, \textbf{k})v^jG^A(E_f, \textbf{k}) \nonumber\\
&&-(i\leftrightarrow j)],
\end{eqnarray}
and
\begin{eqnarray}\label{eq:ext_2}
\sigma_{ij}^\textnormal{II}&=&\frac{e^2\hbar}{2\pi}\int\frac{\textnormal{d}^3k}{(2\pi)^3}\int^{E_f}_{-\infty}{\rm Re}\lbrace{\rm Tr}[\boldsymbol{\Gamma}^i(E, \textbf{k})G^R(E, \textbf{k})  \nonumber\\
& &\times \gamma(E, \textbf{k}) G^R(E, \textbf{k})\boldsymbol{\Gamma}^j(E, \textbf{k})G^R(E, \textbf{k})  \nonumber\\
& &-(i\leftrightarrow j)]\rbrace \textnormal{d}E.
\end{eqnarray}
Here, $\gamma(E, \textbf{k})$ and $\boldsymbol{\Gamma}(E, \textbf{k})$ are scalar and vector vertex functions, respectively, that correct for the identity and velocity operators ($I$ and $\boldsymbol{v}$) adapted to the unperturbed Green's functions $G_0$.  The two vertex functions can be calculated iteratively:
\begin{eqnarray}\label{eq:gamma}
\gamma(E, \textbf{k})&=&I+\mathcal{V}\int\frac{\textnormal{d}^3k'}{(2\pi)^3}O_{\textbf{kk}'}G^R(E, \textbf{k}')\gamma(E, \textbf{k}')\nonumber\\
&&\times G^R(E,\textbf{k}')O_{\textbf{k}'\textbf{k}},
\end{eqnarray}
\begin{eqnarray}\label{eq:Gamma}
	\boldsymbol{\Gamma}(E, \textbf{k})&=&\boldsymbol{v}(\textbf{k})+\mathcal{V}\int\frac{\textnormal{d}^3k'}{(2\pi)^3}O_{\textbf{kk}'}G^A(E, \textbf{k}')\boldsymbol{\Gamma}(E, \textbf{k}') \nonumber\\
	&&\times G^R(E, \textbf{k}')O_{\textbf{k}'\textbf{k}}.
\end{eqnarray}
In the GD model, the skew scattering term mainly comes from the vertex corrections (Eqs.~\eqref{eq:Gamma} and~\eqref{eq:gamma}) and converges to a finite value in  the clean limit \textcolor{blue}{($\mathcal{V}\rightarrow0$)}.  It is therefore called ``intrinsic" skew scattering which differs from the conventional skew scattering~\cite{Sinitsyn2008,Sinitsyn2007}.  By subtracting the intrinsic ($\sigma^\textnormal{int}_{ij}$) and intrinsic skew scattering ($\sigma^\textnormal{isk}_{ij}$) terms from the total AHC ($\sigma^\textnormal{tot}_{ij}$), the side jump term ($\sigma^\textnormal{sj}_{ij}$) can be then obtained.

Similarly to the AHE, the ANE and ATHE also have an intrinsic and extrinsic origin.  The total AHC, ANC, and ATHC can be thus decomposed into three distinct terms
\begin{eqnarray}\label{eq:ANE_ATHE}
	\sigma_{ij}^\textnormal{tot}&=&\sigma_{ij}^\textnormal{int}+\sigma_{ij}^\textnormal{sj}+\sigma_{ij}^\textnormal{isk}, \\
	\alpha_{ij}^\textnormal{tot}&=&\alpha_{ij}^\textnormal{int}+\alpha_{ij}^\textnormal{sj}+\alpha_{ij}^\textnormal{isk}, \\
	\kappa_{ij}^\textnormal{tot}&=&\kappa_{ij}^\textnormal{int}+\kappa_{ij}^\textnormal{sj}+\kappa_{ij}^\textnormal{isk}.
\end{eqnarray}
By plugging the decomposed AHC into Eq.~\eqref{eq:LB}, one can then obtain the corresponding components of ANC and ATHC.

The electronic structure calculations are performed by the full-potential linearized augmented plane-wave (FP-LAPW) method as implemented in the \textsc{FLEUR} code~\cite{fleur}.  The generalized gradient approximation (GGA) with the Perdew-Burke-Ernzerhof (PBE) parameterization~\cite{Perdew1996} is used to treat the exchange-correlation functional.  Spin-orbit coupling is included in all calculations.  The plane-wave cutoff energies of 3.50 $a_0^{-1}$ and 3.40 $a_0^{-1}$ are adopted for Fe$_2$CoAl and Fe$_2$NiAl, respectively.  A uniform $k$-mesh of $12\times12\times12$ was used for self-consistent calculations.  After obtaining the converged ground state charge densities, the maximally localized Wannier functions were constructed from the $s$-, $p$-, and $d$-orbitals of Fe and Co (or Ni) atoms as well as the $s$-orbital of Al atom on a uniform $k$-mesh of $8\times8\times8$ using the \textsc{Wannier90} package~\cite{Pizzi2020}.  After that, anomalous electric, thermal, and thermoelectric transport properties were calculated using the  \textit{ab initio} tight-binding Hamiltonian in the basis of Wannier functions.  To converge the AHC, an ultra-dense $k$-mesh of $350\times350\times350$ was used.  To calculate the ANC and ATHC (refer to Eq.~\eqref{eq:LB}), the AHC is initially calculated with an energy interval of 0.02 eV and then is interpolated to 0.1 meV.

\section{Results and discussion}\label{results}
\begin{figure}
	\includegraphics[width=0.9\columnwidth]{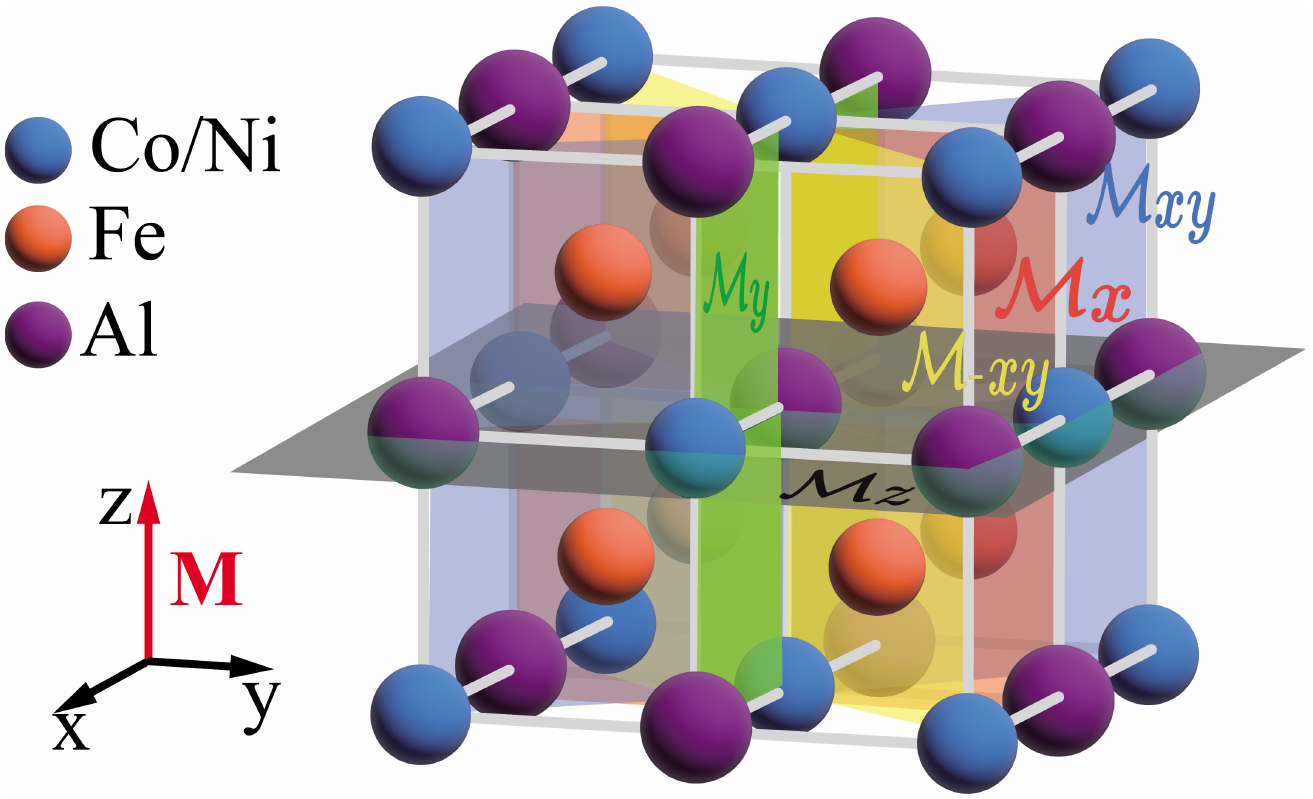}
	\caption{The structure of L2$_1$ ordered cubic Heusler compounds Fe$_2$CoAl and Fe$_2$NiAl.  The spin magnetic moments on Fe, Co, and Ni atoms are along the [001] direction ($z$-axis).  Except for the mirror plane $\mathcal{M}_z$, the system contains other four mirror planes $\mathcal{M}_x$, $\mathcal{M}_y$, $\mathcal{M}_{xy}$, and $\mathcal{M}_{-xy}$ that should be combined  with the time-reversal symmetry $\mathcal{T}$.}
	\label{fig:structure}
\end{figure}

The Heusler compounds Fe$_2$CoAl (FCA) and Fe$_2$NiAl (FNA) have complex crystal structures as they can crystallize in ordered structures (L2$_1$ or X phases) as well as in disordered structures (A2 or B2 phases)~\cite{AHMAD2021,Ahmad2020EPJB,Felix2021}.  Since the ordered L2$_1$ phases of FCA and FNA have been successfully prepared in a recent experimental work~\cite{Saito2018}, we here investigate the magnetic and anomalous transport properties of FCA and FNA with the L2$_1$ phases.  Figure~\ref{fig:structure} shows the crystal structure of cubic L2$_1$ phase, which belongs to the crystallographic space group Fm$\bar{3}$m (No. 225).  The Fe, Co (Ni), and Al atoms occupy the 8c, 4a, and 4b Wyckoff positions, respectively.  The relaxed lattice constants of FCA and FNA are 5.73 {\AA} and 5.74 {\AA}, respectively, in a good agreement with the experimental data (5.732 {\AA} for FCA~\cite{Vishal2013} and 5.758 {\AA} for FNA~\cite{Buschow1983}).  Both FCA and FNA are ferromagnetic metals with ultra-high Curie temperatures ($830\sim1010$ K)~\cite{Saito2018,AHMAD2021}, and hence potentially provide an excellent material platform for spintronics.  The calculated spin magnetic moments of Fe (Fe) and Co (Ni) atoms in FCA (FNA) are 2.08 $\mu_B$ (2.04 $\mu_B$) and 1.82 $\mu_B$ (0.58 $\mu_B$), respectively.

Before performing calculations, we first determine the symmetry properties of the conductivity tensor, especially its nonvanishing off-diagonal elements, using magnetic group theory~\cite{Zhou2019,ZhouXD2020,XD-Zhou2019b,RW-Zhang2021}. Since the symmetry requirements on AHC, ANC, and ATHC are the same according to Eq.~(\ref{eq:LB}), in the following we take the AHC as an example.  The anomalous Hall vector ($\bm{\sigma}$) can be regarded as a pseudovector, like spin, so its vector-form notation, $\boldsymbol{\sigma}$ = [$\sigma^x$, $\sigma^y$, $\sigma^z$] = [$\sigma_{yz}$, $\sigma_{zx}$, $\sigma_{xy}$], is used here for convenience.  Utilizing the \textsc{Isotropy} software~\cite{isotropy}, the magnetic space (point) group for both FCA and FNA is $I4/mm'm'$ ($4/mm'm'$) when the magnetization direction is along one of the three orthogonal crystal axes.  Considering the symmetry of fcc lattice, we only need to discuss the case of $z$-axis magnetization.  
%Due to the translationally invariant of $\bm{\sigma}$, 
Moreover, it is sufficient to restrict our analysis to the magnetic point group, $4/mm'm'$, which contains one mirror plane $\mathcal{M}_z$ and four combined symmetries $\mathcal{TM}_x$, $\mathcal{TM}_y$, $\mathcal{TM}_{xy}$, and $\mathcal{TM}_{-xy}$ (here, $\mathcal{T}$ is the time-reversal symmetry).  The mirror plane $\mathcal{M}_z$ is perpendicular to the magnetization, while the mirror planes $\mathcal{M}_x$, $\mathcal{M}_y$, $\mathcal{M}_{xy}$, and $\mathcal{M}_{-xy}$ are parallel to the magnetization, as shown in Fig.~\ref{fig:structure}.  The mirror symmetry $\mathcal{M}_z$ reverses the sign of $\sigma^x$ and $\sigma^y$ but preserves $\sigma^z$ such that only $\sigma^z$ is nonzero.  The time-reversal symmetry $\mathcal{T}$ reverses the sign of $\sigma^x$, $\sigma^y$, and $\sigma^z$ but the combined symmetries $\mathcal{TM}$ have to reverse $\sigma^z$ again, resulting in nonvanishing $\sigma^z$.  Overall, the magnetic point group $4/mm'm'$ results in an anomalous Hall vector $\bm{\sigma}$ = [0, 0, $\sigma^z$].  Therefore it is sufficient to calculate the $z$-component of anomalous transport properties only, that is, $\sigma^z(=\sigma_{xy})$, $\alpha^z(=\alpha_{xy})$, and $\kappa^z(=\kappa_{xy})$.

\begin{figure}[t]
	\includegraphics[width=\columnwidth]{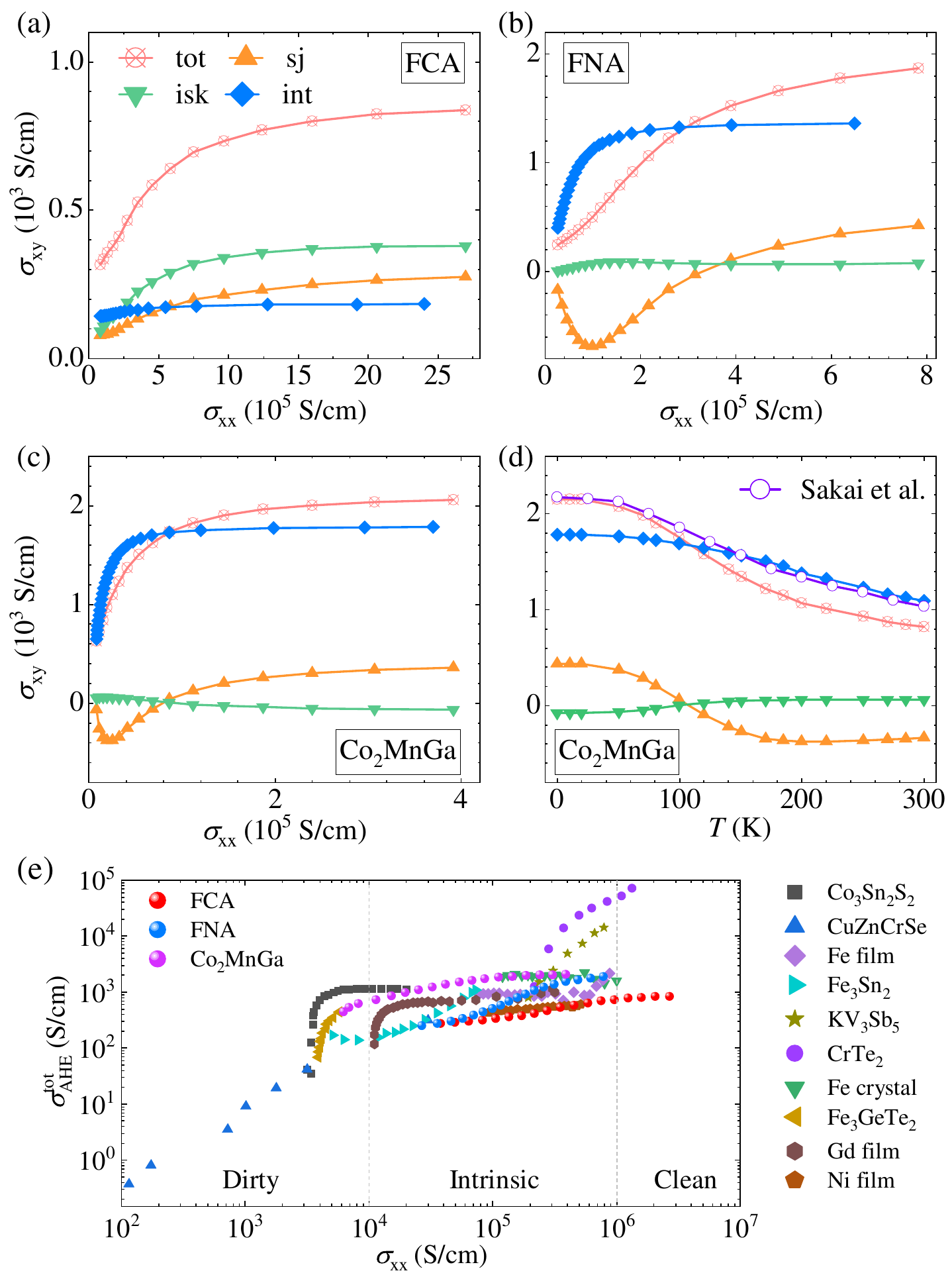}
	\caption{Intrinsic versus intrinsic contributions to the anomalous transport. (a-c) The total AHC ($\sigma_{xy}^\textnormal{tot}$) and its decomposition (intrinsic $\sigma_{xy}^\textnormal{int}$, side-jump $\sigma_{xy}^\textnormal{sj}$, and intrinsic skew-scattering $\sigma_{xy}^\textnormal{isk}$) as a function of longitudinal conductivity ($\sigma_{xx}$) for FCA, FNA, and Co$_2$MnGa.  (d) The temperature-dependent AHC for Co$_2$MnGa presented in comparison to experimental data  by Sakai et al~\cite{Akito2018}.  (e) AHC $\sigma_{xy}^\textnormal{tot}$ versus $\sigma_{xx}$ for FCA, FNA, and Co$_2$MnGa ranging across intrinsic from dirty to clean regimes.  The data for other magnetic materials are taken from Refs.~\cite{Miyasato2007,YangSY2020_2,HuangM2021}.}
	\label{fig:ahc}
\end{figure}

In order to reveal the dominant mechanism of the AHE in FCA and FNA, the total AHC ($\sigma_{xy}^\textnormal{tot}$) and its decomposition ($\sigma_{xy}^\textnormal{int}$, $\sigma_{xy}^\textnormal{sj}$, and $\sigma_{xy}^\textnormal{isk}$) are plotted as a function of the longitudinal conductivity $\sigma_{xx}$ in Figs.~\ref{fig:ahc}(a) and~\ref{fig:ahc}(b), respectively.  It is clear that the extrinsic contributions ($\sigma_{xy}^\textnormal{sj}+\sigma_{xy}^\textnormal{isk}$) play a dominant role in FCA except for the case of extremely low longitudinal conductivity $\sigma_{xx}<2\times10^5$ S/cm. For $\sigma_{xx}$ larger than $20\times10^5$ S/cm, the extrinsic contributions account for nearly $78\%$ of the total AHC.  On the contrary, the intrinsic mechanism dominates the AHE in FNA overall,  contributing by about $\sim87\%$ to the total AHC when $\sigma_{xx}>6\times10^5$ S/cm.  In the clean limit, the total AHC $\sigma_{xy}^\textnormal{tot}$ converges to the saturated values of 837 S/cm and 1868 S/cm for FCA and FNA, respectively.  The large AHC of FNA is comparable to that of famous Co$_2$MnGa, the AHC for which is plotted for comparison in Fig.~\ref{fig:ahc}(c).  The calculated total AHC $\sigma_{xy}^\textnormal{tot}$ for Co$_2$MnGa is 2060 S/cm, where the intrinsic part $\sigma_{xy}^\textnormal{int}$ is 1759 S/cm and the extrinsic side-jump (skew scattering) part $\sigma_{xy}^\textnormal{sj}$ ($\sigma_{xy}^\textnormal{isk}$) is 361 ($-$60) S/cm.  A good agreement of our calculated total AHC with the experimental value ($\sim$2000 S/cm~\cite{Akito2018}) demonstrates the importance of including the extrinsic contributions into consideration when comparing theory with experiment.

The skew scattering is found to contribute mostly in FCA, while it can be ignored in FNA and Co$_2$MnGa.  Another prominent feature is the non-monotonic variation (first decreasing and then increasing) of the side-jump term as $\sigma_{xx}$ increases, which appears in both FNA and Co$_2$MnGa but not in FCA.  To gain a deeper understanding of this anomalous behavior, further analysis is needed to examine the competition between the three universality classes of side-jump scattering (spin-independent, spin-conserving, and spin-flip) in relation to the spin structures of the disorder potential~\cite{SYAYang2011}.  The GD model we use here generally accounts for all of the mean-field scattering channels, whereas the details of scattering sources are not explicitly specified.  Employing the experimental data of Co$_2$MnGa measured by Sakai et al.~\cite{Akito2018}, the temperature-dependence of AHC is mapped out, as shown in Fig.~\ref{fig:ahc}(d), from which one can see that our calculations agree well with the experiments.  We thus expect that the results for FCA and FNA will be confirmed by future experiments.  A common feature among the three Heusler compounds is that the AHC gradually increases with the increasing of longitudinal conductivity $-$ a trend which has also been observed in other magnetic materials, as depicted in Fig.~\ref{fig:ahc}(e).

Next, from the analysis of the electronic structure we provide arguments as to why the extrinsic and intrinsic mechanisms dominate the AHE in FCA and FNA, respectively.  First of all, for a variety of magnetic materials, different scaling relations have been proposed by analyzing the dependency of $\sigma_{xy}$ on $\sigma_{xx}$~\cite{Nagaosa2010,Miyasato2007,YangSY2020_2,HuangM2021}: $\sigma_{xy} \propto \sigma_{xx}^{1.6}$ in the dirty regime ($\sigma_{xx} < 10^4$ S/cm), nearly constant $\sigma_{xy}$ in the intrinsic regime ($10^4 < \sigma_{xx} < 10^6$ S/cm), and $\sigma_{xy} \propto \sigma_{xx}^{2}$ or $\sigma_{xy} \propto \sigma_{xx}^{1}$ in the clean regime ($\sigma_{xx} > 10^6$ S/cm).  From Fig.~\ref{fig:ahc}(e), one can directly understand the reason for the specific origin of the AHE in our  Heusler compounds: while $\sigma_{xx}$ of FNA falls into the intrinsic regime where intrinsic AHE dominates,  $\sigma_{xx}$ of FCA is large enough to reach into the clean regime, where the origin of the  AHE is expected to be extrinsic~\cite{Onoda2006}. In this context, the case of Co$_2$MnGa is similar to that of FNA.

\begin{figure}[t]
	\includegraphics[width=1\columnwidth]{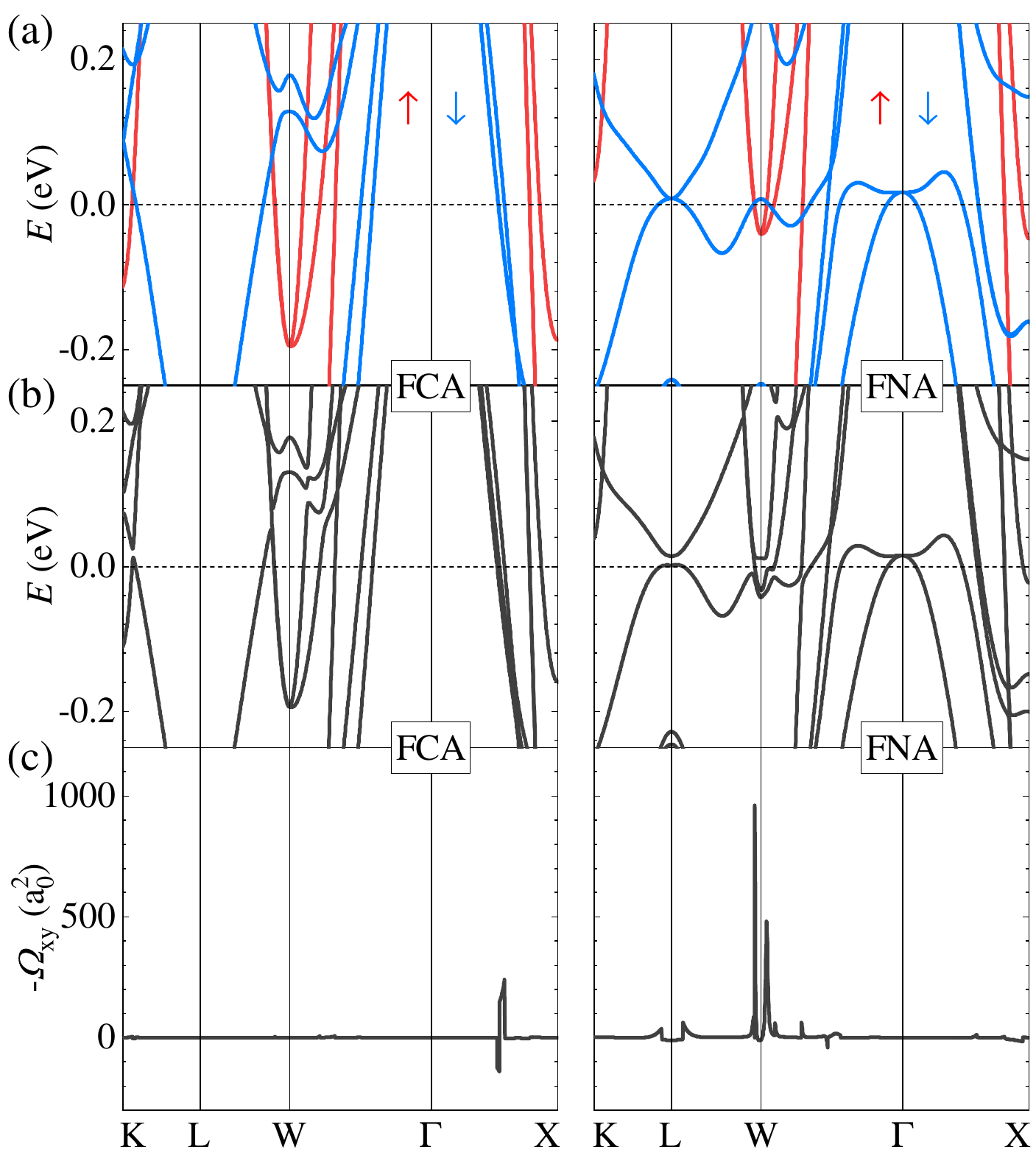}
	\caption{Band structures and Berry curvature. (a,b) Band structures without (a) and with (b) spin-orbit coupling for FCA (left) and FNA (right). In (a), spin-up and spin-down bands are marked with red and blue lines, respectively. (c) Corresponding Berry curvature $\Omega_{xy}$ along high-symmetry lines.}
	\label{fig:band}
\end{figure}

The intrinsic AHC $\sigma_{xy}^\textnormal{int}$ of FCA is nearly one order of magnitude smaller than that of FNA (see Table~\ref{tab:atp}), which can be understood from band structure and Berry curvature analysis.  In Figs.~\ref{fig:band}(a) and~\ref{fig:band}(b), we plot the band structures of FCA and FNA without and with spin-orbit coupling, respectively.  Both FCA and FNA are ferromagnetic metals, in which the spin-up and spin-down bands cross the Fermi energy ($E_f$) individually.  The band crossings between opposite spin channels in FCA are away from $E_f$, while the ones in FNA locate exactly at the Fermi energy, for example, at the W point.  After turning on the spin-orbit coupling, the band crossings at W  are gaped out, giving rise to a large Berry curvature ($\Omega_{xy}$) in FNA (see Fig.~\ref{fig:band}(c)).  The Berry curvature in FCA is peaked along the $\Gamma$--X path, however, the positive and negative peaks mostly cancel each other.  We thus obtain a much larger $\sigma_{xy}^\textnormal{int}$ in FNA than in FCA.  Moreover, steep band dispersion in FCA results in a small electron's effective mass and small density of states, producing large $\sigma_{xx}$.  On the other hand, the nearly flat bands in FNA near  $E_f$ around L, W, and $\Gamma$  result in a large electron's effective mass and large density of states, giving rise to a small $\sigma_{xx}$, in agreement to explicit calculations (Fig.~\ref{fig:ahc}(e)).  Additionally, the L2$_1$ phases of FCA and FNA exhibit a larger $\sigma_{xy}$ than the corresponding X phases or some disordered phases~\cite{Felix2021}, similar to the case of Co$_2$MnAl~\cite{Sakuraba2020}.

\begin{figure*}[t]
	\includegraphics[width=1.8\columnwidth]{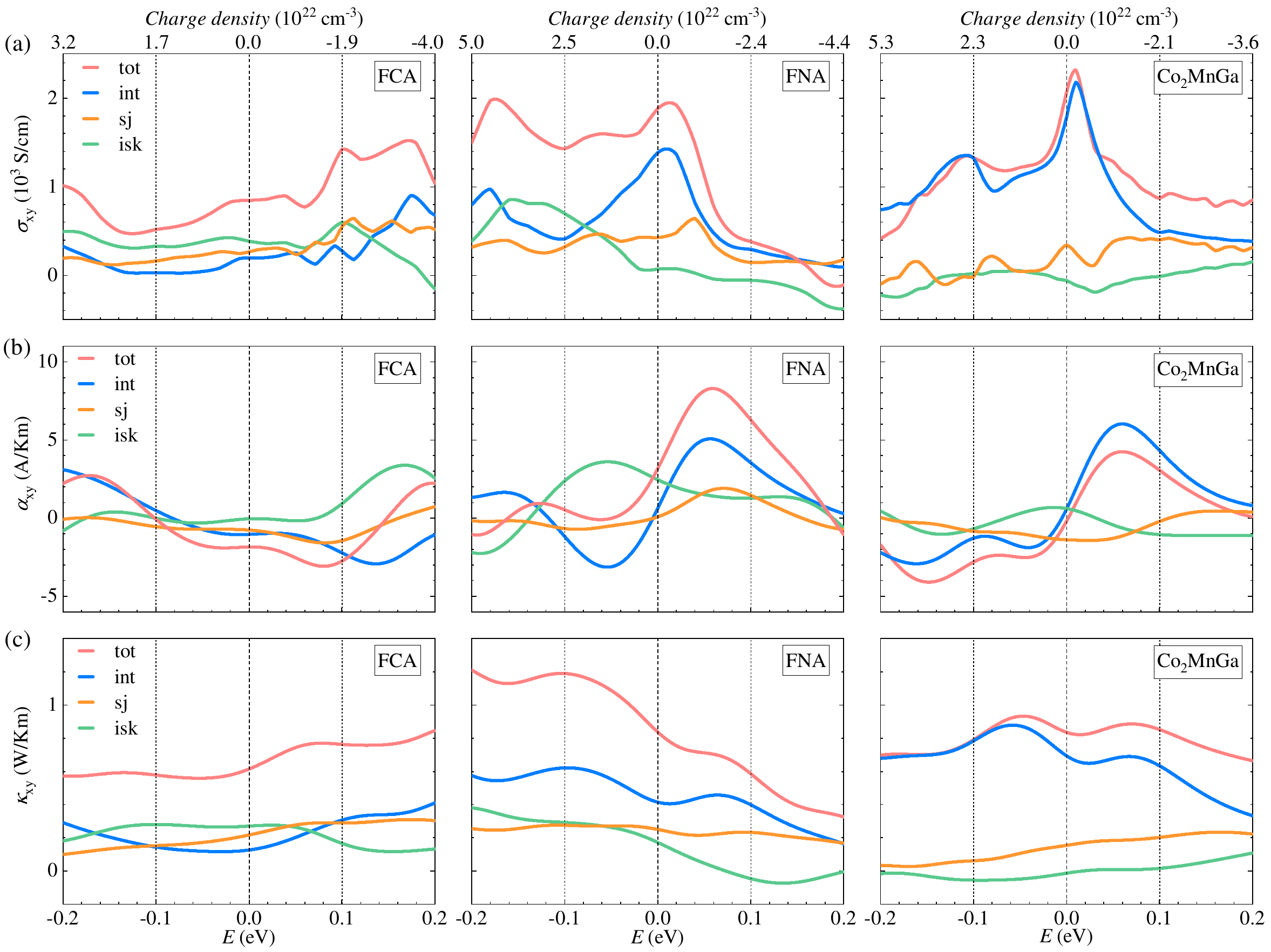}
	\caption{ The total and decomposed AHC $\sigma_{xy}$ (a), ANC $\alpha_{xy}$ (b), and ATHC $\kappa_{xy}$ (c) calculated in the clean limit as a function of Fermi energy for FCA (left), FNA (middle), and Co$_2$MnGa (right).  The upper axes correspond to the values of charge doping concentrations.  The ANC and ATHC are calculated at the temperature of 300 K.}
	\label{fig:atp_E}
\end{figure*}

\begin{figure}[t]
	\includegraphics[width=1\columnwidth]{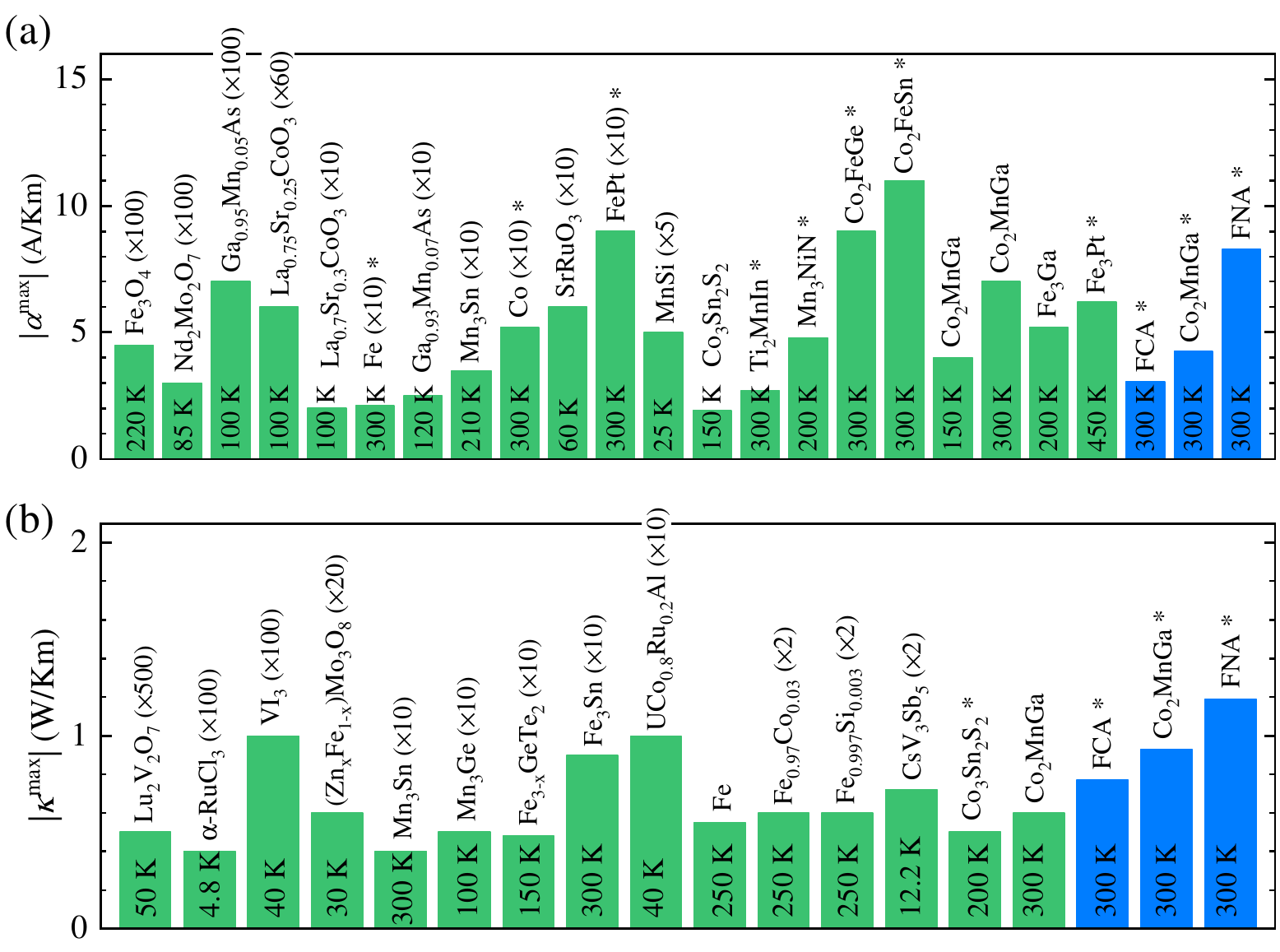}
	\caption{The maximal values of the ANC (a) and ATHC (b) for FCA, FNA, and Co$_2$MnGa computed here  (in blue) are presented  in comparison to other typical magnets reported in previous works~\cite{Akito2018,ZhouXD2020,Akito2020,LiXK2017,XuLC2020,Miyasato2007,ANE_Jonathan2018,ZhangHD2021,ZhangH2021,Yokoi_RuCl3_2021,ZhouXB2022,Guin2019,Tdeue_multiferroics_2017,Sugii2019,Shiomi_ATHE_2009,Karmaker2022,ANE_RamosFe3O4_2014,ANE_HanasakiNd2Mo2O7_2008,ANE_YongGaMnAs_2008,ANE_Weischenberg2013,ANE_MuhammadMn3Sn_2017,ANE_HirokaneMnSi_2016,ANE_Noky2018,ANE_GuinCo3Sn2S2_2019,ATHE_Onose_2010,ATHE_Taishi_2021,ATHE_Asaba_2021,ATHE_Macy_2021} (green color).  The numbers in parentheses correspond to prefactors used to scale the original data.  The asterisks mark computational results.}
	\label{fig:anc_athc}
\end{figure}

\begin{table}[t]\footnotesize
	\caption{The total AHC, ANC, and ATHC as well as their decompositions at the true Fermi energy and under appropriate charge doping for FCA, FNA, and Co$_2$MnGa.  The units of AHC, ANC, and ATHC are S/cm, A/Km, and W/Km, respectively.  The corresponding charge doping concentrations can be read from Fig.~\ref{fig:atp_E}.}
	\label{tab:atp}
	\begin{ruledtabular}
		\begingroup
		\setlength{\tabcolsep}{4.5pt} % Default value: 6pt
		\renewcommand{\arraystretch}{1.5} % Default value: 1
		\begin{tabular}{crrcrrcrr}
			&\multicolumn{2}{c}{FCA} &  &\multicolumn{2}{c}{FNA} &  &\multicolumn{2}{c}{Co$_2$MnGa}\\
			\cline{2-9}
			$E_f$ (eV) & 0  &+0.10  & &0  &+0.01   &  &0  &+0.01\\
			\hline
			$\sigma_{xy}^\textnormal{int}$   &183   &276   &  &1368  &1421   & &1759   &2178 \\          
			$\sigma_{xy}^\textnormal{sj}$    &275   &548   &  &423   &451    & &361    &251 \\	 
			$\sigma_{xy}^\textnormal{isk}$   &379   &598   &  &77    &75     & &-60    &-111  \\
			$\sigma_{xy}^\textnormal{tot}$   &837   &1422  &  &1868  &1947   & &2060   &2318 \\
			\\
			\cline{2-9}
			$E_f$ (eV) & 0  &+0.08  & &0  &+0.06   &  &0  &+0.06\\
			\hline                          
			$\alpha_{xy}^\textnormal{int}$   &-1.04   &-1.66   &  &0.63  &5.06   & &0.58   &6.01 \\          
			$\alpha_{xy}^\textnormal{sj}$    &-0.76   &-1.58   &  &0.08  &1.81    & &-1.39    &-1.17 \\	 
			$\alpha_{xy}^\textnormal{isk}$   &-0.05   &0.16   &  &2.46    &1.42     & &0.61    &-0.60  \\
			$\alpha_{xy}^\textnormal{tot}$   &-1.85   &-3.08  &  &3.17  &8.29   & &-0.20   &4.24 \\
			\\
			\cline{2-9}
			$E_f$ (eV) & 0  &+0.07  & &0  &-0.10   &  &0  &-0.05\\
			\hline                          
			$\kappa_{xy}^\textnormal{int}$   &0.13   &0.25   &  &0.41  &0.62   & &0.69   &0.87 \\          
			$\kappa_{xy}^\textnormal{sj}$    &0.22   &0.30   &  &0.25   &0.28    & &0.15    &0.11 \\	 
			$\kappa_{xy}^\textnormal{isk}$   &0.27   &0.22   &  &0.17    &0.29     & &-0.01    &-0.05  \\
			$\kappa_{xy}^\textnormal{tot}$   &0.62   &0.77  &  &0.83  &1.19   & &0.83   &0.93 \\
		\end{tabular}
		\endgroup
	\end{ruledtabular}
\end{table}  

To evaluate the effect of charge doping on the anomalous transport properties of FCA and FNA, we plot the total AHC, ANC, and ATHC as well as their decompositions obtained in the clean limit  as a function of Fermi energy  in Fig.~\ref{fig:atp_E}, together  with the values for  Co$_2$MnGa.  The energy range is varied from $-$0.2 eV to $+$0.2 eV, which corresponds to an achievable charge doping concentration on the order of $10^{22}$ cm$^{-3}$ (see upper axes in Figs.~\ref{fig:atp_E} and corresponding vertical dotted lines).  We find that the magnitude of anomalous transport properties can be tuned quite significantly by doping. For example, the total AHC $\sigma_{xy}^\textnormal{tot}$ increases up to 1422 S/cm, 1947 S/cm, and 2318 S/cm for FCA, FNA, and Co$_2$MnGa by shifting the Fermi energy upward by 0.10 eV, 0.01 eV, and 0.01 eV, respectively.  Another prominent feature is that the proportions of intrinsic ($\sigma_{xy}^\textnormal{int}$), side jump ($\sigma_{xy}^\textnormal{sj}$), and skew scattering ($\sigma_{xy}^\textnormal{isk}$) terms can change with varying of Fermi energy, i.e. with doping.  For example in case of the AHC in FCA the dominant mechanism switches from extrinsic to intrinsic when $E>$ 0.15 eV (left panel of Figs.~\ref{fig:atp_E}(a)), while in FNA the skew scattering dominates in the energy range of $-1.6\sim-0.08$ eV instead of the leading intrinic contribution  elsewhere in energy (middle panel of Figs.~\ref{fig:atp_E}(a)).  Interestingly, the intrinsic mechanism dominates the AHC in Co$_2$MnG in the overall energy range (right panel of Figs.~\ref{fig:atp_E}(a)).  The total and decomposed AHC for FCA, FNA, and Co$_2$MnGa at the true Fermi energy and under appropriate charge doping are summarized in Tab.~\ref{tab:atp}.

The total and decomposed ANC, i.e., $\alpha_{xy}^\textnormal{tot}$, $\alpha_{xy}^\textnormal{int}$, $\alpha_{xy}^\textnormal{sj}$, and $\alpha_{xy}^\textnormal{isk}$, in FCA, FNA, and Co$_2$MnGa, calculated at 300 K, is plotted in Fig.~\ref{fig:atp_E}(b).  At the $E_f$, $\alpha_{xy}^\textnormal{tot}$ is $-$1.85 A/Km in FCA and 3.17 A/Km in FNA, with both values being much larger than in Co$_2$MnGa ($-$0.20 A/Km).  It should be noted that by tuning the $E_f$ upward by 0.06 eV,  $\alpha_{xy}^\textnormal{tot}$ in Co$_2$MnGa increases up to 4.24 A/Km, agreeing well with the experimental values of $\sim 4$ A/Km (150 K)~\cite{Akito2018} and $\sim 7$ A/Km (300 K)~\cite{Guin2019}.  Similarly, $\alpha_{xy}^\textnormal{tot}$ in FCA reaches as much as $-$3.08 A/Km at $E_f+0.08$ eV, and  $\alpha_{xy}^\textnormal{tot}$ in FNA is peaked at $E_f+0.06$ eV with a colossal value of 8.29 A/Km, exceeding most of current magnetic materials, as shown in Fig.~\ref{fig:anc_athc}(a).  The competition between intrinsic and extrinsic mechanisms for the ANC can also be seen in the three Heusler compounds.  For example, the intrinsic and extrinsic contributions dominate the ANC  in FCA under large hole ($E<-0.1$ eV) and electron ($E>0.1$ eV) dopings, respectively, while they are similar in magnitude around the Fermi energy.  In the cases of FNA and Co$_2$MnGa, a crossover from intrinsic to extrinsic mechanism occurs by switching the electron doping ($E>0$) to hole doping ($E<0$).

Additionally, Fig.~\ref{fig:atp_E}(c) displays remarkable ATHE properties at 300 K for FCA, FNA, and Co$_2$MnGa.  The calculated $\kappa_{xy}^\textnormal{tot}$ of FCA and FNA are 0.62 W/Km and 0.83 W/Km at the $E_f$, respectively.  The calculated $\kappa_{xy}^\textnormal{tot}$ of Co$_2$MnGa is 0.83 W/Km that is comparable with the experimental data ($\sim$ 0.6 W/Km at 300 K)~\cite{XuLC2020}.  Similar to the AHE and ANE, doping electrons or holes can greatly enhance the ATHE.  For example, $\kappa_{xy}^\textnormal{tot}$ can be increased up to 0.77 W/Km at 0.07 eV for FCA, 1.19 W/Km at $-$0.10 eV for FNA, and 0.93 W/Km at $-$0.05 eV for Co$_2$MnGa, respectively.  These values are much larger than most of typical magnetic materials, see Fig.~\ref{fig:anc_athc}(b).  The calculated ATHC values for FCA, FNA, and Co$_2$MnGa at the true Fermi energy and under appropriate charge doping are collected in Tab.~\ref{tab:atp}.

\section{Summary}
In conclusion, utilizing first-principle density functional theory calculations, we systematically investigated the intrinsic and extrinsic AHE, ANE, and ATHE in ferromagnetic Heusler compounds Fe$_2$CoAl and Fe$_2$NiAl.  In the clean limit, the intrinsic mechanism dominates the AHE in Fe$_2$NiAl.  It can be understood from the nearly flat bands around the Fermi energy, which leads to a relatively large electron's effective mass and large density of states such that the longitudinal conductivity falls into the intrinsic regime.  In contrast, the extrinsic mechanism dominates the AHE in Fe$_2$CoAl.  The physics can be explained from the steep band dispersion across the Fermi energy, which produces a small electron's effective mass and a high longitudinal conductivity falling into the clean regime.  The larger intrinsic contribution in Fe$_2$NiAl can also be understood from the strong Berry curvature around the band crossings gapped out by spin-orbit coupling, features which are lacking  in Fe$_2$CoAl.  Another famous Heusler ferromagnet Co$_2$MnGa is studied for comparison, where we find the origin of the AHE to be identical to Fe$_2$NiAl.  Similarly to the AHE, the intrinsic mechanism of the ATHE dominates in both Fe$_2$NiAl and Co$_2$MnGa, while the extrinsic mechanism plays a dominant role in Fe$_2$CoAl. In the case of the ANE, the intrinsic and extrinsic mechanisms contribute with similar magnitude at the true Fermi energy for the three Heusler compounds.  By introducing appropriate electron or hole doping, the proportions of the intrinsic and extrinsic contributions as well as the total values of AHE, ANE, and ATHE can be effectively tuned.  In particular, the ANC and ATHC in Fe$_2$NiAl can reach up to the recorded values of 8.29 A/Km and 1.19 W/Km respectively, which are much larger than most of the current ferromagnetic and antiferromagnetic materials.  Our results show that Fe$_2$CoAl and Fe$_2$NiAl host excellent anomalous transport properties and therefore provide a promising material platform for spintronics and spin-caloritronics.

\begin{acknowledgments}
This work is supported by the National Key R\&D Program of China (Grant Nos. 2022YFA1402600, 2022YFA1403800, and 2020YFA0308800), the National Natural Science Foundation of China (Grant Nos. 12274027, 11874085, 12274028, and 52161135108), the Science \& Technology Innovation Program of Beijing Institute of Technology (Grant No. 2021CX01020).   Y.M. acknowledges the Deutsche Forschungsgemeinschaft (DFG, German Research Foundation) TRR 288 - 422213477 (project B06). Y.M., W.F., and Y.Y. acknowledge the funding under the Joint Sino-German Research Projects (Chinese Grant No. 12061131002 and German Grant No. 1731/10-1) and the Sino-German Mobility Programme (Grant No. M-0142).
\end{acknowledgments}

\bibliography{./references.bib}

\end{document}